\documentclass[a4paper,11pt,reqno]{article}
\usepackage[body={16cm,24cm}]{geometry}

\usepackage{amssymb}  
\usepackage{amsmath}  
\usepackage{latexsym} 
\usepackage{graphicx}
\usepackage{hyperref}
\usepackage{booktabs}
\usepackage{adjustbox}
\usepackage{aas_macros}
\newcommand*{\email}[1]{%
    \normalsize\href{mailto:#1}{#1}\par
}

\begin{document}

\title{\textbf{The plausibility of origins scenarios requiring two impactors}}

\author{R.~J. Anslow$^{1*}$, A. Bonsor$^{1}$, P.~B. Rimmer$^{2}$, A.~S.~P. Rae$^{3}$, C.~H. McDonald$^{1}$ and\\ C.~R. Walton$^{4,1}$}
\date{%
    {\small
    $^1$Institute of Astronomy, University of Cambridge, Madingley Road, Cambridge, CB3 0HA, UK\\%
    $^2$Astrophysics Group, Cavendish Laboratory, University of Cambridge, JJ Thomson Ave, Cambridge, CB3 0HE, UK\\%
    $^3$Department of Earth Sciences, University of Cambridge, Cambridge CB2 3EQ, UK\\
    $^4$Department of Earth Sciences, Institute f{\"u}r Geochemie und Petrologie, ETH Z{\"u}rich, Zurich, Switzerland\\
    $^\ast$Corresponding author: \email{rja92@ast.cam.ac.uk}
    }
}

\maketitle
\vspace{-3em}
%%%% Abstract text to be placed here %%%%%%%%%%%%
\begin{abstract}
    Hydrogen cyanide delivered by cometary impactors can be concentrated as ferrocyanide salts, which may support the initial stages of prebiotic chemistry on the early Earth. One way to achieve the conditions required for a variety of prebiotic scenarios, requiring for example the formation of cyanamide and cyanoacetylene, is through the arrival of a secondary impactor. In this work, we consider the bombardment of the early Earth, and quantitatively evaluate the likelihood of origins scenarios that invoke double impacts. Such scenarios are found to be possible only at very early times ($>\,$4\,Gya), and are extremely unlikely settings for the initial stages of prebiotic chemistry, unless (i) ferrocyanide salts are stable on 1000\,yr timescales in crater environments, (ii) there was a particularly high impact rate on the Hadean Earth, and (iii) environmental conditions on the Hadean Earth were conducive to successful cometary delivery (i.e., limited oceanic coverage, and low ($\lesssim 1$\,bar) atmospheric surface pressure). Whilst environmental conditions on the early Earth remain subject to debate, this work highlights the need to measure the typical lifetime of ferrocyanide salts in geochemically realistic environments, which will determine the plausibility of double impact scenarios.
\end{abstract}
%%%%%%%%%%%%%%%

\medskip


\section{Introduction}
\label{sec:introduction}

The surfaces of planetary bodies in our Solar System bear the accumulated scars of hypervelocity collisions -- impact craters. The spatial distribution, individual sizes and depth, and morphology of craters encodes information about the rate, and nature of impact events over time. This information is of great relevance for understanding the evolution of planetary bodies, from the collisional accretionary processes of their birth, to the environmentally constructive and/or disruptive consequences of later stage events.

Much attention has been paid to the flux of impactors to the early Earth in the context of life's origins. It is thought that early massive impactors (i.e., $\gtrsim 10$\,-$100\,$km) may have repeatedly sterilised the surface of putative life \cite{MaherStevenson1988, AbramovMojzsis2009, Abramov2013}, yet may simultaneously have created transient reducing conditions that fostered prebiotic chemistry \cite{Benner2019, Zahnle2020, Itcovitz2022}. These massive, as well as smaller, impactors may also have delivered an important complement of siderophile \cite{Mann2012, Rubie2015}, and volatile elements to Earth's surface (e.g., carbon, nitrogen; \cite{Marty2012, Halliday2013, Marty2016}), furnishing important bulk stockpiles of elements necessary for prebiotic chemistry \cite{ChybaSagan1992, Pearce2017, Osinski2020, Walton2024}. Impacts may also play an important role through altering geological environments on local scales \cite{Pastorek2019}, and selectively catalysing the formation of important prebiotic molecules \cite{Rotelli2016}. Less attention, however, has been paid to the spatiotemporal patterns of cratering, which critically determine whether impactors may have plausibly modified environmental conditions and chemical feedstocks on the right timescales, and in the right locations to foster prebiotic chemistry. 

Impacts have been invoked in a variety of prebiotic chemical scenarios, not simply as sources of molecules, but as instigators of molecular transformation, see for example \cite{Ferus2017a, Ferus2020}. The cyanosulfidic scenario for the origin of life invokes impacts for both delivery and transformation \cite{Patel2015, Sasselov2020}. The first impact could deliver soluble phosphorus in the form of schreibersite (which occurs in both asteroids and comets \cite{Brownlee2012}), and could also transform nitrogen into cyanide \cite{Ferus2017b}. Cyanide will, however, undergo hydrolysis in aqueous solution, which may limit its utility in prebiotic chemistry depending on the pH and temperature of its environment \cite{Miyakawa2002}. One way to circumvent this problem is to store this cyanide in a more stable form, which would allow for more efficient prebiotic synthetic chemistry at a later point. Ferrocyanide is more stable than cyanide \cite{Todd2024}, is itself useful in prebiotic chemistry \cite{Xu2018}, and will form quickly given a source of ferrous iron (Fe(II)) and hydrogen cyanide (HCN) \cite{KeefeMiller1996, Sasselov2020}. The formation of ferrocyanide is therefore, potentially, a key intermediate step required for HCN to be a productive feedstock for prebiotic chemistry, and may form readily on the early Earth \cite{KeefeMiller1996, Sasselov2020}.

Recent observations have demonstrated a large diversity in CHN- and CHS-bearing molecules on the comet 67P \cite{Altwegg2017}, which is also reflected in the abundance of HCN observed in Solar System comets \cite{MummaCharnley2011}. The intact cometary delivery of HCN, the simplest N-bearing organic molecule, has therefore emerged as an additional, atmosphere-independent source of HCN in high (local) concentrations \cite{PierazzoChyba1999, Todd2020, Anslow2023}. In addition, comets are the only empirically-grounded source of prebiotic concentrations of cyanoacetylene \cite{MummaCharnley2011}; other sources either imply steady-state surface concentrations orders of magnitude lower than that used in prebiotic experiments\cite{Wogan2023}, or lack experimental support \cite{ShorttleRimmer2024}. In this study we therefore consider a cometary impact to be the initial source of HCN, which can rain out into water rich with Fe(II) to form ferrocyanide. There are, however, several alternative sources of HCN on the early Earth. For instance, the synthesis of HCN (and even more complex organic molecules) is possible via photochemistry in transiently reducing post-impact atmospheres \cite{Zahnle2020, Wogan2023}, via impact shocks in reducing atmospheres \cite{Ferus2017a, Parkos2018}, or during high-velocity impacts themselves \cite{Goldman2010, Cassone2018}. The formation of ferrocyanide would then still be possible, given an environmental source of Fe(II), and would no longer necessarily be confined to the physical location of an impact crater.

Once this water dries, depending on the leftover cations \cite{TonerCatling2019}, the surviving ferrocyanide precipitate \cite{Todd2022} could then be thermally transformed via a second impact, into iron or magnesium cyanamide, potassium cyanide and calcium carbide \cite{Patel2015}. These salts are presumed to be stable long-term when dry, and once rehydrated, would release hydrogen cyanide, cyanamide and cyanoacetylene, and the schreibersite would supply phosphate \cite{Bryant2013}. These salts therefore store, segregate and release the very feedstock molecules required for the production of amino acids, nucleotides and phospholipids \cite{Patel2015,Green2021}. These same molecules are invoked in several other prebiotic chemical scenarios \cite{OroKimball1961, Ferris1968, Lohrmann1968, Benner2012, Parker2014, Becker2019}. Though other sources of heat, such as volcanism or lightning, may be sufficient to generate these salts \cite{Sasselov2020}, these sources risk equilibration with more neutral to oxidizing media, namely the magma or the atmosphere. It is currently unclear, however, how the redox state of the magma or atmosphere would affect the formation of ferrocyanide salts.

The ability of multiple impactors to supply (via the delivery of HCN), and interact with stockpiles of ferrocyanide is dependent on the typical lifetime of ferrocyanide on the early Earth. Aqueous ferrocyanide is known to decompose in UV light \cite{Asperger1952}, which is potentially significant in the context of the enhanced UV-flux on the early Earth \cite{RanjanSasselov2017}. Todd et al.\ \cite{Todd2022} investigated this in more detail, and found that millimolar (mM) concentrations of ferrocyanide would have a very short half-life, on the order of hours. Whilst these results favour ferrocyanide survival in (alkaline) carbonate lake environments, as proposed in \cite{TonerCatling2019}, in the absence of a plausible UV-shielding mechanism it is very likely ferrocyanide will degrade on prohibitively short timescales \cite{Todd2022}. One potential way to protect ferrocyanide from UV degradation is when precipitated as salts \cite{Sasselov2020}, however to the best of our knowledge the typical lifetime of ferrocyanide salts in these conditions has not been measured. This timescale will constrain the number of environments able to productively use ferrocyanide on the early Earth.

In this work we consider the bombardment of the early Earth, and evaluate the time- and spatial-scales over which materials were delivered, and environments were perturbed. In doing so, we quantitatively evaluate the likelihood of scenarios for the origin of life that invoke {\it double impacts}, i.e., the production of overlapping craters suitable for prebiotic chemistry. We define double impact scenarios as consisting of an initial cometary impact, in which HCN is successfully delivered (leading to the formation of ferrocyanide salts), and a subsequent smaller impact responsible for the thermal metamorphosis of ferrocyanide salts.

This paper starts in \S\ref{sec:methods} by introducing a Monte Carlo crater overlap model and meteoroid atmospheric entry model, which we use to quantitatively evaluate the frequency of double impact scenarios on the early Earth. In \S\ref{sec:results}, we present the results of these numerical simulations, which include the minimum cometary diameter able to survive atmospheric entry, the corresponding crater diameter, and from this the number of overlapping craters on the early Earth as a function of time. In \S\ref{sec:discussion} we discuss the feasibility of double impact scenarios, given our numerical results, and comment on plausible alternative routes for prebiotic chemistry. Finally, \S\ref{sec:conclusions} summarises our conclusions.

\section{Methods}
\label{sec:methods}

In this study we aim to assess the feasibility of double impact scenarios on the early Earth, which is directly linked to the probability of two impactors occurring in the same location (i.e., the production of overlapping craters). This probability is estimated using the Monte Carlo model described in \S\ref{sec:methods_montecarlo}. To accurately assess the feasibility of double impact scenarios, we must relate the distribution of impactors at the top of the Earth's atmosphere to the resulting impact crater distribution, which is dependent on the composition of the impactor population, the corresponding dynamics of atmospheric entry, and the physics of crater formation. A numerical model describing the atmospheric entry of comets, and the scaling relations used for corresponding impact crater formation are introduced in \S\ref{sec:methods_atmos}, and \S\ref{sec:methods_cratering} respectively.
In \S\ref{sec:methods_chronology}, we discuss both the Earth's impact chronology, and constraints on the successful delivery of HCN, from which we are ultimately able to quantify the feasibility of double impact scenarios.

\subsection{A Monte Carlo crater overlap model}
\label{sec:methods_montecarlo}
We use a Monte Carlo approach to estimate the number of overlapping craters on a planetary surface. Specifically, our model answers the question ``What is the fraction, $f_{\rm overlap}$, of overlapping craters we expect to find within a population of $N$ craters distributed randomly over the surface of a planet?''.

We first calculate the probability that two craters, randomly distributed over a sphere, intersect.
To determine whether two craters intersect, we calculate the distance between the centre of the two craters along the surface of the sphere. The location of the centre of these craters are defined by the vectors $\mathbf{x}_1$ and $\mathbf{x}_2$, relative to the centre of the sphere, which allow us to calculate the distance between the two craters as
\begin{equation}
    d\left(\mathbf{x}_1, \mathbf{x}_2\right) = R_\oplus \arccos{\left(\frac{\mathbf{x}_1\cdot\mathbf{x}_2}{R_\oplus^2}\right)},
\end{equation}
where $R_\oplus$ is the radius of the Earth. Overlapping craters are identified when $d\left(\mathbf{x}_1, \mathbf{x}_2\right) < R_1 + R_2$, where $R_1$ and $R_2$ are the radii of the two craters, which are drawn from a power-law size frequency distribution (SFD) \cite{Bland2005, HergartenKenkmann2015}. We enforce the additional constraint that $R_2 < R_1$, with the aim of ensuring the secondary impactor does not totally erase any extant chemistry in the initial crater.

In our Monte Carlo code, following a similar approach to \cite{overlap_montecarlo}, we first calculate the probability, $p_{\rm single}$, that two craters randomly placed on the surface of a sphere do not overlap. To do this, we generate an initial population of $N_{\rm MC}=10^6$ craters, each with a random location, and radius drawn from a power-law SFD. For each crater we then perform an additional $N_{\rm MC}$ trials in which we generate a new crater (with random radius and location), and check for overlap as described above. By counting the number of overlapping craters we are able to estimate the probability, $p_i$, that the second crater does not overlap the initial crater. The probability, $p_{\rm single}$, that these two randomly drawn craters do not overlap is therefore 
\begin{equation}
    p_{\rm single} = \frac{1}{N_{\rm MC}}\sum\limits_{i=1}^{N_{\rm MC}}{p_i}.
\end{equation}
For a population of $N$ craters, the probability that none of the remaining craters will overlap with the initial crater is $p_{\rm single}^{N-1}$, since each pair of craters is independent, and we average over the crater's SFD when calculating $p_{\rm single}$. The expected number of overlapping craters in the population is then $N\left(1 -  p_{\rm single}^{N-1}\right)$, however in this work we typically consider the fraction of overlapping craters within the population, given by
\begin{equation}
    f_{\rm overlap} = 1 - p_{\rm single}^{N-1},
\end{equation}
such that for small $N$, $f_{\rm overlap} \sim 0$, corresponding to no overlapping craters, and for large $N$, $f_{\rm overlap} \sim 1$, corresponding to a geometrically saturated surface (e.g., \cite{Hartmann1984, Melosh1989}) on which all craters will have at least one intersection with another crater.

The SFD of terrestrial craters is an important parameter in this model; shallower distributions, with a greater proportion of large craters, will saturate a surface with fewer impactors. There are however significant complexities associated with determining the true SFD of the terrestrial crater population, as it is incomplete at small sizes \cite{GrieveShoemaker1994}, with the atmosphere obscuring the flux of smaller impactors, and geological activity rapidly erasing the record of smaller craters \cite{BlandArtemieva2006, HergartenKenkmann2015}. Recent work has suggested the terrestrial crater record is complete for $D \gtrsim 6\,$km \cite{HergartenKenkmann2015}, and we make the assumption this same distribution continues down to smaller sizes. We therefore assume a cumulative SFD given by
\begin{equation}
    n(D) \propto D^{-2},
\end{equation}
matching the terrestrial record for larger craters \cite{GrievePesonen1992, Hughes2000}. This will likely overestimate the true number of sub-kilometre craters, as the atmosphere is able to filter out these impactors (which is not true for the Moon). By modelling the interaction of comets with the atmosphere, we are better able to characterise the formation of small craters on the Earth, and estimate a minimum cometary diameter capable of surviving atmospheric entry, which we discuss next. We discuss later in \S\ref{sec:discussion_limitations} the sensitivity of our results to the exact value of the power-law index.

\subsection{Cometary atmospheric entry model}
\label{sec:methods_atmos}

A meteoroid entering the atmosphere is subject to a number of competing processes through which it loses kinetic energy; through its velocity, due to atmospheric drag, and its mass, due to ablation (e.g., \cite{BaldwinSheaffer1971, Bronshten1983}). The differential pressure across the meteoroid increases with atmospheric density, causing the deformation and fragmentation of the body as this pressure exceeds its material strength. The dispersal of these fragments greatly increases the meteoroid's cross-sectional area, further increasing its energy loss. In what can be a relatively rapid process, small meteoroids therefore deposit the majority of their initial kinetic energy in an explosive airburst, which occurs within a fraction of an atmospheric scale height (e.g., \cite{HillsGoda1993, Chyba1993}). 
Small (i.e. sub-km) comets are particularly vulnerable to fragmentation, and will typically airburst at high altitude, a consequence of their structural fragility. Large comets however, are able to escape this fate, since pressure waves do not have sufficient time to propagate through the comet before impact \cite{Chyba1990}. The size and velocity of comets arriving at the top of the atmosphere may therefore be significantly different to those arriving at the Earth's surface, with significant consequences for the diameter of the subsequent impact crater.

To understand, qualitatively, which comets are most sensitive to the stresses of atmospheric entry we follow a similar argument to Melosh \cite{Melosh1989}. As a comet enters the atmosphere, it will encounter a column-integrated atmospheric mass $P_{\rm surf}/g\sin{\theta}$ per unit area (where $P_{\rm surf}$ is the surface atmospheric pressure, $g$ the gravitational acceleration, and $\theta$ the angle with respect to the local horizontal). By the conservation of momentum, we expect that atmospheric deceleration will begin to play a significant role only when the total atmospheric mass it interacts with equals the comet's mass. This critical comet size is therefore,
\begin{equation}
	\label{eq:r_crit_qualitative}
	r_{\rm crit} \sim \frac{3}{4} \frac{P_{\rm surf}}{\rho_mg \sin{\theta} },
\end{equation}
where $\rho_m$ is the comet's density. For Earth's current atmosphere, this simple calculation suggests atmospheric entry will be very challenging for comets smaller than 10-100\,m (depending on the comet's bulk density), and that this value will increase with decreasing cometary bulk density, and increasing atmospheric surface density.

The detailed modelling of fragmentation is however very challenging, with various approaches including progressive fragmentation models (in which individual fragments are tracked; e.g., \cite{PasseyMelosh1980, SvetsovNemtchinov1995}), deformation/pancake models (in which fragment material behaves like a fluid and spreads laterally; e.g., \cite{Zahnle1992, Chyba1993, FieldFerrara1995}), and 2/3D hydrodynamical models (which, at great numerical expense, attempt to consider all effects; e.g., \cite{Korycansky2000, Shuvalov2013}). In the interests of simplicity, we opt to use the model from Chyba et al.\ (1993) \cite{Chyba1993}, which works well for Tunguska-like impactors, and should provide an accurate picture of the basics of cometary atmospheric entry.
The equations describing the drag, and ablation of a meteoroid are given by 
\begin{align}
    \label{eq:comet_dvdt}
    m\frac{dv}{dt} &= -\frac{1}{2}C_D\rho_aAv^2 + mg\sin{\theta},\\
    Q\frac{dm}{dt} &= -A \cdot {\rm min}\left(\frac{1}{2}C_H\rho_av^3, \sigma_{\rm SB} T^4\right),
\end{align}
where $A$ is the meteoroid's cross-sectional area, $m$ its mass, and $v$ its velocity. The atmospheric density is $\rho_a$, and the gravitational acceleration is given by $g=GM_\oplus/(R_\oplus + h)^2$, where $h$ is the meteoroid's altitude. The two coefficients, $C_D$ and $C_H$, describe the drag and heat-transfer efficiencies respectively, whilst $Q\,[{\rm J\,kg^{-1}}]$ is the meteoroid's heat of ablation. 
The ablation equation is based on the classic evaporative ablation model of Sekanina \cite{Sekanina1993}, whilst also including an upper limit, proportional to $\sigma_{\rm SB} T^4$, dictated by the efficiency of thermal ionisation, where $T\approx25,000\,$K is the temperature of the shocked gas in front of the meteoroid. The trajectory of the meteoroid is described by
\begin{align}
    \frac{d\theta}{dt} &= \frac{g\cos{\theta}}{v} - \frac{C_L\rho_aAv}{2m} - \frac{v\cos{\theta}}{R_\oplus + h}\\
    \frac{dh}{dt} &= -v\sin{\theta},
\end{align}
where $C_L$ is the meteoroid's lift coefficient, which relates the rate of change of $\theta$ to the differential force across the meteoroid. For simplicity we assume an isothermal atmospheric profile,
\begin{equation}
    \rho_a = \rho_0 \exp{\left(-\frac{h}{H}\right)},
\end{equation}
where $\rho_0$ and $H$ are the atmospheric surface density, and scale height respectively.

As discussed, the meteoroid must withstand a large ram pressure when passing through the atmosphere if it is to survive intact, with the leading-face experiencing a pressure ($P_s = C_D \rho_a v^2 / 2$) much larger than anything experienced by the trailing face. When the average pressure experienced by the meteoroid ($P_S /2$) exceeds its tensile strength, it is assumed the fragment material will flow laterally, therefore increasing its cross-sectional area. The expansion of the meteoroid's radius, $r$, is described by
\begin{equation}
    \label{eq:comet_deformation}
    r\frac{d^2r}{dt^2} = \frac{C_D\rho_av^2}{2\rho_m},
\end{equation}
where $v$ is the meteoroid's velocity, and $\rho_m$ its density. As before, $C_D$ is the drag coefficient, and $\rho_a$ the atmospheric density. 
This is, however, an idealised picture of fragmentation and in practice there will be some critical radius after which fragments will develop their own individual bow shocks \cite{PasseyMelosh1980, HillsGoda1993}. In any case, the surface area of the meteoroid will significantly increase, causing the rapid deceleration and the deposition of energy into the atmosphere in an airburst-type event. Following Chyba et al.\ \cite{Chyba1993} and Collins et al.\ \cite{Collins2005}, we define this this to be when the radius reaches 6 times its initial value, ensuring good agreement with Tunguska-class events.

\subsection{Cometary impact crater formation}
\label{sec:methods_cratering}
As discussed in the previous section, small impactors undergo complex interactions with the atmosphere, affecting their impact velocity, diameter, and in extreme cases causing them to airburst at high altitude. This is most pronounced for comets due to their low density, and tensile strength, and therefore directly determines a minimum possible crater size from cometary impacts. This quantity, when combined with the Earth's impact chronology, directly determines the typical distribution of cometary craters on the Earth.

The formation of an impact crater depends on the physical properties of both the target and impactor, as well as the impact velocity and angle. Here, we use the scaling law from Schmidt \& Housen \cite{SchmidtHousen1987} which relates the diameter of the impactor, $D_{\rm imp}$, to the transient crater, $D_{\rm tr}$, as
\begin{equation}
    D_{\rm tr} = 1.161 \left(\frac{\rho_{\rm imp}}{\rho_{\rm tar}}\right)^{1/3}D_{\rm imp}^{0.78}v_{\rm imp}^{0.44}g^{-0.22}\sin^{1/3}{\left(\theta\right)}
\end{equation}
where $\rho_{\rm tar}$ ($\rho_{\rm imp}$) is the density of the target (impactor), $g$ the acceleration due to gravity of the target, $v_{\rm imp}$ the impact velocity, and $\theta$ the impact angle with respect to the surface normal. 

Following Korycansky \& Zahnle \cite{KorycanskyZahnle2005}, we re-cast this scaling relation in terms of the mass, diameter, and kinetic energy of the impactor to account for the non-spherical geometry of impactors as a result of fragmentation and deformation, which acts to increase the diameter of an impactor at roughly constant density. The assumption here is that fragmented impactors will be less effective at cratering, having already imparted a significant fraction of their initial kinetic energy to the atmosphere. Detailed numerical modelling would however be required to accurately model the physical process of crater formation, which is beyond the scope of this work. In terms of the impactor's kinetic energy, the transient crater diameter is given by
\begin{equation}
    D_{\rm tr} = 1.677 \, M_{\rm imp}^{0.113} \, D_{\rm imp}^{-0.22} \, \rho_{\rm tar}^{-{1/3}} \, \left(\frac{M_{\rm imp} v_{\rm imp}^2}{2}\right)^{0.22} \, g^{-0.22} \, \sin^{1/3}{\left(\theta\right)}.
\end{equation}

For larger impactors the transient crater is gravitationally unstable, which leads to the formation of a complex crater. Following Croft (1985) \cite{Croft1985}, a further-scaling law is used to apply this correction to the transient crater diameter, 
\begin{equation}
    D_{\rm crater} = \begin{cases}
        1.25 \, D_{\rm tr} \hspace{1em} &{\rm if} \hspace{1em} D_{\rm tr} < D_c,\\
        1.17 \, D_{\rm tr}^{1.13}\,D_c^{-0.13} \hspace{1em} &{\rm if} \hspace{1em} D_{\rm tr} > D_c,
    \end{cases}
\end{equation}
where $D_c$ is the simple-complex crater transition diameter, which scales inversely with surface gravity \cite{Pike1980}. For the Earth we assume a value of 4\,km for $D_c$ \cite{Ivanov2001}.

\subsection{Calculating the number of overlapping craters on the early Earth}
\label{sec:methods_chronology}

Understanding how the impact rate on the Earth has declined since the Moon-forming impact is crucially important in determining the number of overlapping craters on the early Earth as a function of time. The number of these craters capable of supporting subsequent prebiotic chemistry is dependent on the proportion of cometary impacts, and the fraction of these impacts in which prebiotically relevant concentrations of prebiotic feedstock molecules will be delivered. We will now motivate the choices made in this work, with the assumed values summarised in Table~\ref{tab:model_assumptions}.

\subsubsection{The proportion of cometary impacts on the early Earth}
\label{sec:proportion_cometary_impactors}

We assume that comets contribute approx.\ 1\% of the total impacts, in light of several lines of evidence for a predominantly asteroidal impactor flux on the early Earth. Firstly, the lunar crater size distribution is consistent with a population of main asteroid belt (MAB) impactors, with a SFD similar to that of near-Earth asteroids \cite{Strom2005, Richardson2009}. The next line of evidence comes from the analysis of lunar melts, and meteorite fragments in lunar regolith breccias, which suggest a dominant asteroidal contribution to the early lunar bombardment \cite{KringCohen2002, Joy2012}. Finally, a low ($\lesssim\,$1\%) cometary contribution to Earth's volatile inventory is recorded by both water D/H ratios \cite{Dauphas2000, Morbidelli2000}, and several noble gas isotope ratios \cite{Marty2016}. This picture has since been supported by dynamical modelling (e.g., \cite{Bottke2012, Rickman2017, Nesvorny2023, Joiret2023, Joiret2024}), and is consistent with the current impact flux on the Earth, which is dominated by near-Earth asteroids \cite{Weissman2007, YeomansChamberlin2013}. It is unlikely the fraction of cometary impacts was independent of time, however given the lack of direct evidence constraining this fraction, and the rough agreement at early and late times, we argue this is a reasonable assumption to make. The influence of this rough estimate on our results is discussed in detail in \S\ref{sec:discussion_limitations}.

\subsubsection{Constraints on successful cometary delivery}
\label{sec:proportion_cometary_delivery}

The successful delivery of prebiotic molecules is very challenging at high impact velocities \cite{Todd2020, Anslow2023}, with peak temperatures (experienced by the leading edge of the comet) briefly reaching 5,000\,-\,10,000\,K \cite{PierazzoChyba1999}. Furthermore, successful cometary delivery will require the presence of subaerial continental crust in order for there to be localised aqueous environments in which these molecules can be plausibly concentrated; the global delivery of HCN is unlikely to be relevant, with the oceans ultimately remaining too dilute \cite{Todd2020}.
The typical velocity distribution of cometary impactors, and the extent of oceanic coverage on the early Earth will therefore dictate the frequency of successful cometary delivery.

Numerical simulations of terrestrial planet formation record mean impact velocities in excess of 20\,km\,s$^{-1}$ for comets, asteroids, and leftover planetesimals (e.g., \cite{Bottke2012, Nesvorn2017, Brasser2020}), which is much faster than the Earth's escape velocity. This is roughly consistent with mean impact velocities of approx.\ 20\,km\,s$^{-1}$ for near-Earth asteroids \cite{StuartBinzel2004, Greenstreet2012}, and 22\,km\,s$^{-1}$ for short-period comets \cite{HughesWilliams2000, Brasser2020}. The successful cometary delivery of prebiotically relevant concentrations of HCN requires oblique, low velocity, small radius impacts \cite{Todd2020}. Moreover, new results suggest that peak temperatures may have been previously underestimated (McDonald et al., in prep.), emphasising the requirement for very low velocity impacts. Assuming a log-normal distribution with mean velocity 22\,km\,s$^{-1}$ and standard deviation $\sim\,$5km\,s$^{-1}$ \cite{Brasser2020}, we find that only 1\% of comets will have impact velocity below 15\,km\,s$^{-1}$. It is likely only these comets will deliver sufficient concentrations of HCN for subsequent prebiotic synthesis \cite{Todd2020}. We therefore assume, perhaps optimistically, that 1\% of cometary impacts will successfully deliver prebiotic molecules, but note this fraction is poorly constrained.

There is direct evidence for the presence of oceans during the early Hadean (as early as 4.3-4.4\,Gyr; \cite{Wilde2001, Mojzsis2001}), which would greatly decrease the surface area available for crater formation, necessary for the concentration of HCN following cometary impacts. The extent of oceanic coverage is however subject to debate (e.g., \cite{Korenaga2021}), with the potential for a deep Hadean ocean, a result of efficient degassing of water due to rapid plate tectonics. This would restrict subaerial landmass to hot-spot volcanic islands \cite{BadaKorenaga2018, RosasKorenaga2021}. Several studies assume a linear growth in the fraction of exposed continental crust during the Hadean (e.g., \cite{McCullochBennett1993}), however given the uncertainties in these models we conservatively assume a constant subaerial landmass of 20\% (following Guo \& Korenaga \cite{GuoKorenaga2020}), which is an upper limit in the case of shallow oceans.

\subsubsection{Connecting our model to the terrestrial impact chronology}

For double impact scenarios to successfully reactivate prebiotic chemistry, the secondary impact must occur within some timespan, as determined by the typical lifetime of ferrocyanide salts in crater-pond environments, $\tau_{\rm Fe(CN)_6}$. In the presence of UV-irradiation, we know that ferrocyanide has a very short lifetime in aqueous solution \cite{Todd2022}. The lifetime of ferrocyanide salts could however be significantly longer, and so we leave $\tau_{\rm Fe(CN)_6}$ as a free parameter in our model. There is therefore a competition between the incident impact rate, $R_{\rm in}$, and the degradation of ferrocyanide salts, which will determine the number of craters on the Earth's surface available for subsequent prebiotic chemistry. We model the balance between crater formation and ferrocyanide degradation using a simple sink-source equation,
\begin{equation}
    \frac{dN_{\rm crater}}{dt} = R_{\rm in} - \frac{N_{\rm crater}}{\tau_{\rm Fe(CN)_6}},
\end{equation}
where $N_{\rm crater}$ is the number of craters capable of supporting subsequent prebiotic chemistry. In steady-state the number of craters available for subsequent prebiotic chemistry will be $R_{\rm in}\tau_{\rm Fe(CN)_6}$, which will, when combined with the fraction of successful cometary impacts, be used in our Monte Carlo overlap model (see \S\ref{sec:methods_montecarlo}) to calculate the number of overlapping craters as a function of time. 
\begin{table}[t!]
    \centering
    \begin{adjustbox}{width=\textwidth}
    \begin{tabular}{cc} \toprule
        Model Parameter & Assumed Value \\ \midrule
        The proportion of cometary impactors  & 1\%\,\textsuperscript{(e.g., \cite{KringCohen2002, Nesvorny2023})} \\
        The proportion of successful cometary impacts (i.e., the survival of HCN)  & 1\%\,\textsuperscript{(e.g., \cite{PierazzoChyba1999, Todd2020})} \\
        The fraction of subaerial landmass on the Archean and Hadean Earth  & 20\% (time-independent)\textsuperscript{\cite{GuoKorenaga2020}}\\
        \bottomrule
    \end{tabular}
    \end{adjustbox}
    \caption{Summary of the model assumptions made in this work when calculating the number of overlapping craters on the early Earth. The number of overlapping craters is proportional to the product of these parameters, allowing for the direct comparison of our results to each choice. See \S\ref{sec:methods_chronology} for more detail.}
    \label{tab:model_assumptions}
\end{table}

\begin{figure}[t!]
    \centering
    \includegraphics[width=0.75\textwidth]{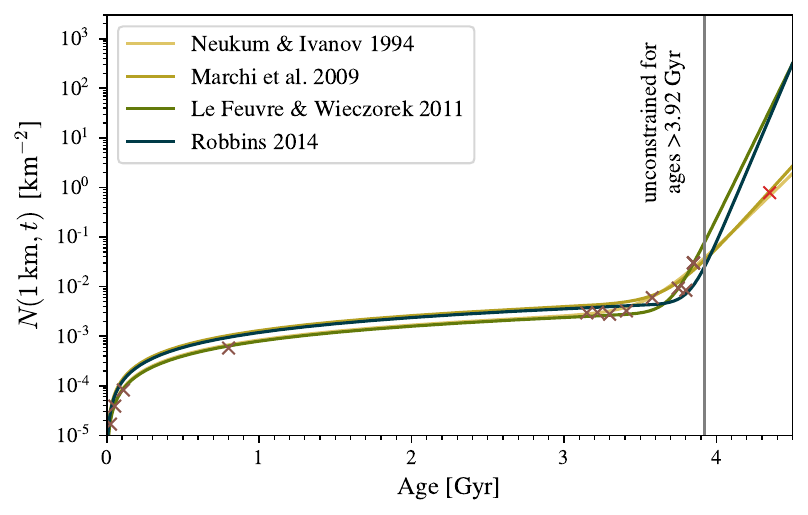}
    \caption{The number of accumulated $D>1\,$km lunar craters, per square kilometre, is plotted as a function of age for a number of proposed impact chronologies. Calibration points (brown crosses) from Le Feuvre \& Wieczorek (2011) \cite{LeFeuvreWieczorek2011} are included for reference. The inclusion of the (disputed) Highland crust, as indicated by the red cross, differentiates between the fits of Neukum \& Ivanov (1994), Marchi (2009) \cite{NeukumIvanov1994, Marchi2009} and Le Feuvre \& Wieczorek (2011), Robbins (2014) \cite{LeFeuvreWieczorek2011, Robbins2014}, highlighting the inherent uncertainty associated with extrapolating any chronology beyond approx.\ $3.92\,$Gyr. }
    \label{fig:crater_chronology_methods_1km}
\end{figure}

The time-dependence of the incident impact rate is crucially important in determining when, and if double impact scenarios could have been possible on the early Earth.  Despite the incomplete terrestrial crater record, the distribution, and morphology of impact craters on the surface of the Moon allows us to model the chronology of impacts in the inner solar system \cite{NeukumIvanov1994, Neukum2001, Hartmann2001}. 
There is clear evidence for a high early impact rate, in the form of a collection of approx.\ 3.9\,Gyr lunar samples \cite{Tera1974, Ryder1990}, which initially motivated the impact spike hypothesis of the Late Heavy Bombardment (LHB) \cite{Gomes2005}. Recent studies however suggest these impact ages may instead reflect a sampling bias, and instead record a more prolonged bombardment, which has decayed monotonically since 4.5\,Gyr \cite{Zellner2017}. In addition to this, there is a dearth of pre-3.92\,Gyr samples, due to challenges calibrating absolute ages, and an almost complete lack of radiometric data between 1-3\,Gyr (\cite{StofflerRyder2001}, see also Figure~\ref{fig:crater_chronology_methods_1km}). 
More recently, the age and composition of the samples returned by Chang'e-5 have improved constraints on the inner Solar System impact chronology, supporting lower estimates of the impact rates during the 1-3\,Gyr interval \cite{Che2021}. Nonetheless, it is clear that it remains very difficult to constrain the true impact rate through the 1-3\,Gyr interval, and pre-3.92\,Gyr.

All impact chronologies are therefore an extrapolation between these data, and a commonly adopted approach has been to assume the accumulated density of craters with diameter $D$ after a time $t$, $N(t, D)$, is a separable function of both diameter and time. Different approaches have converged on qualitatively similar impact chronologies, which take the following functional form,
\begin{equation}
    N(t, D) = N(D) \left\{a\left(\exp{(bt)}-1\right) + ct\right\},
\end{equation}
where $N(D)$ is the cumulative crater SFD, and $a$, $b$ and $c$ are constants. 
Four different impact chronologies \cite{NeukumIvanov1994, Marchi2009, LeFeuvreWieczorek2011, Robbins2014} are shown in Figure~\ref{fig:crater_chronology_methods_1km}, in which we see that the absence of agreed upon calibration data after approx.\ $3.92\,$Gyr leads to significant discrepancies between the respective fits. This clearly highlights the difficulty associated with extrapolating impact rates close to the Moon-forming impact.

\section{Results}
\label{sec:results}
\begin{figure}[!t]
    \centering
    \includegraphics[width=0.75 \textwidth]{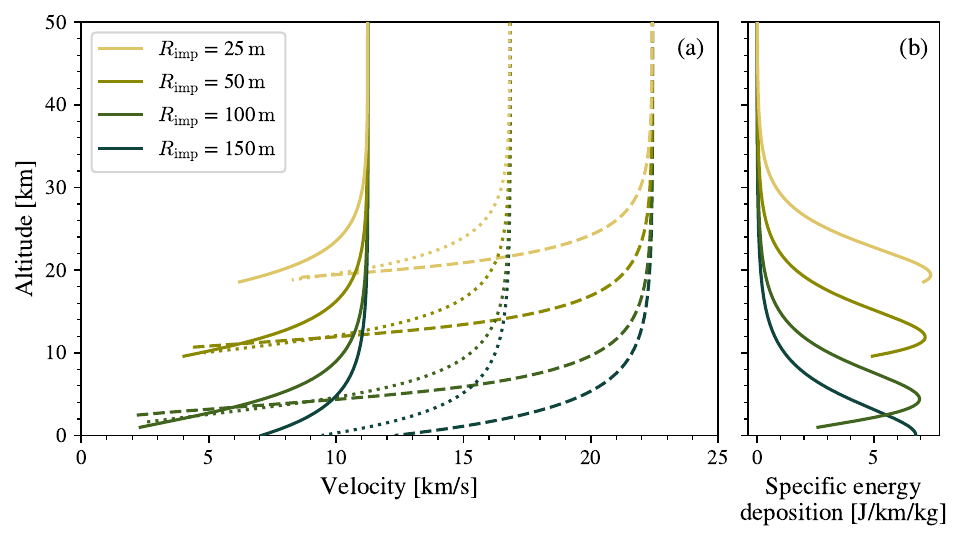}
    \caption{The comets' trajectories are calculated using the described model (see \S\ref{sec:methods_atmos}), assuming an oblique impact at $45^\circ$, a cometary density of $0.6\,{\rm g\,cm}^{-3}$ \cite{LowryWeissman2003, Kokotanekova2017}, a tensile strength of $10^4\,$Pa \cite{Groussin2018}, and an isothermal atmosphere with scale height $7.2\,$km and surface density $1.225\,{\rm kg\,m}^{-3}$ (typical values for the Earth's current atmosphere). (a) The comet's velocity is plotted as a function of altitude for a range of initial velocities and radii. The smallest comets airburst at high altitudes, with $R_0=150\,$m an minimum threshold for cometary survival. (b) The atmospheric kinetic energy deposition is plotted as a function of altitude (for initial velocity $11.2\,{\rm km\,s}^{-1}$). For all but the $150\,$m comet, the kinetic energy is all deposited into the atmosphere in an explosive airburst-type event.}
    \label{fig:comet_survival_trajectories}
\end{figure}
The atmospheric entry model, as described in \S\ref{sec:methods_atmos}, is used to calculate the minimum diameter comet capable of surviving atmospheric entry. We then calculate the corresponding crater diameter, using the comet's diameter and velocity directly before impact. This crater diameter is used in the Monte Carlo overlap model (see \S\ref{sec:methods_montecarlo}) to calculate the number of overlapping craters on the Earth as a function of time.

\subsection{The minimum cometary diameter to reach the Earth's surface}
\label{sec:results_atmos}

Only the largest comets avoid catastrophic fragmentation, and total ablation in the atmosphere, and are able to reach the surface of the Earth. Using the analytical atmospheric entry model of Chyba et al.\ \cite{Chyba1993}, as discussed in \S\ref{sec:methods_atmos}, we vary the initial diameter and velocity of the comet to determine this critical diameter. The dynamics of the comet in the atmosphere are sensitive to both the assumed atmospheric profile, and the comet's physical properties, for which we assume upper limits for both density, and tensile strength (see Table~\ref{tab:impactor_properties}), which will correspond to a lower limit on possible cometary crater diameters. As we will demonstrate in \S\ref{sec:results_montecarlo}, this choice maximises the potential number of overlapping craters on the early Earth, thereby providing the most optimistic picture of double impact scenarios.

We assume an isothermal atmosphere with a scale height and surface density of $7.2\,$km and $1.225\,{\rm g\,cm}^{-2}$ respectively, roughly consistent with the Earth's current atmosphere. We note that atmospheric entry will be possible for much smaller comets if the surface density of the Earth's atmosphere was significantly lower at early times, however geological constraints (particularly during the Hadean) remain subject to debate \cite{CatlingZahnle2020, Charnay2020}. This assumption is however roughly consistent with the Archean atmosphere, with evidence from nitrogen and argon isotopes trapped in quartz, and the size of fossilised raindrops \cite{Marty2013, Som2016} supporting a $\lesssim 1\,$bar surface pressure.

As expected, the atmospheric entry of comets is very sensitive to their initial diameter, with comets larger than approx.\ 1\,km able to reach the surface without significant atmospheric interactions. This is because the duration of entry is shorter than the crossing time of shock-driven pressure waves \cite{Chyba1990}. At the other extreme, for initial diameters below approx.\ 10\,m, comets are totally decelerated and ablated, and do not reach the surface. It is at intermediate diameters where comets undergo complex atmospheric interactions and significant fragmentation, sensitive to the details of the chosen numerical fragmentation model. We plot the trajectories for comets in this size range in Figure~\ref{fig:comet_survival_trajectories}, and find that comets larger than roughly $150\,$m are able to reach the surface, at which point they will form impact craters. As discussed, this number should not be taken as an exact value, given both the approximations made within the numerical model, and the fact that some fortuitous fragments may successfully reach the surface (e.g., \cite{Clark1988}).

\begin{table}[t!]
    \centering
    \begin{tabular}{cccc} \toprule
        Impactor type & Density [g\,cm$^{-3}$] & Tensile strength [Pa] & Heat of ablation [J\,kg$^{-1}$] \\ \midrule
        Comet  & 0.6 \textsuperscript{\cite{LowryWeissman2003},\cite{Kokotanekova2017}} & 10$^4$ \textsuperscript{\cite{Groussin2018}} & $2.5\times10^6$ \textsuperscript{\cite{Chyba1990}} \\
        Stone  & 3.5 \textsuperscript{\cite{BaldwinSheaffer1971}}  & $10^7$ \textsuperscript{\cite{Bronshten1983}} & $8\times10^6$ \textsuperscript{\cite{BaldwinSheaffer1971}} \\
        Iron  & 7.8 \textsuperscript{\cite{BaldwinSheaffer1971}}  & $10^8$ \textsuperscript{\cite{Bronshten1983}}  & $8\times10^6$ \textsuperscript{\cite{BaldwinSheaffer1971}} \\ \bottomrule
    \end{tabular}
    \caption{The choice of parameters used in this work to represent the mechanical properties of each impactor type, as relevant for modelling the physics of atmospheric entry, fragmentation, and impact crater formation.}
    \label{tab:impactor_properties}
\end{table}
\begin{figure}[t!]
    \centering
    \includegraphics[width=0.75 \textwidth]{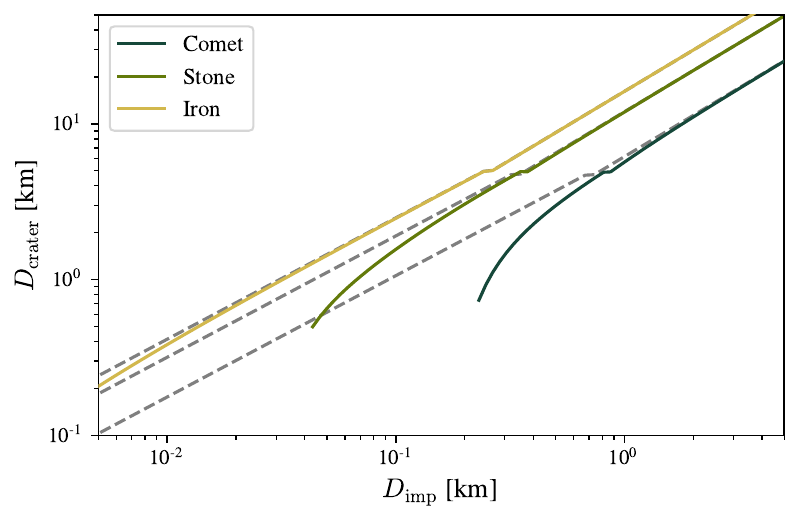}
    \caption{The impact crater diameter is plotted for stony, iron-rich and cometary impactors, as a function of initial diameter (at the top of the atmosphere). The deformation model from Chyba et al.\ \cite{Chyba1993}, as described in \S\ref{sec:methods_atmos}, is used to calculate the deceleration and deformation of the impactors as they pass through the atmosphere. The impactor's final velocity, and diameter at the Earth's surface is used to calculate the diameter of the resulting impact crater, using the scaling laws described in \S\ref{sec:methods_cratering}. For reference, the final crater diameters, not including the effects of the atmosphere, are indicated by the grey dashed lines. The assumed impactor properties are listed in Table~\ref{tab:impactor_properties}.}
    \label{fig:comet_crater_production_constraints}
\end{figure}

\subsection{Results of crater formation}
\label{sec:results_cratering}
We use the crater-scaling laws as introduced in \S\ref{sec:methods_cratering} to calculate the crater size produced by cometary, stony, and iron-rich impactors post-atmospheric entry, using the results of \S\ref{sec:results_atmos} to calculate the impactors' velocity and diameter at the surface. As discussed, the dominant effect of the atmosphere is to filter-out smaller impactors, which is more significant for comets due to their low tensile strength. The results of this calculation can be seen in Figure~\ref{fig:comet_crater_production_constraints}, where we see that the minimum cometary crater diameter is approximately 1\,km, corresponding to roughly 200\,m diameter comets with a initial velocity of $11.19\,{\rm km\,s}^{-1}$ at the top of the atmosphere. As discussed in \S\ref{sec:methods_chronology}, this is not representative of the observed size and velocity distribution of cometary impactors, such that the diameter of typical cometary craters will be significantly in excess of 1\,km.

Stony impactors are typically much stronger, with tensile strengths of approx.\ $5\times10^6\,$Pa, whilst iron meteorites can even exceed $10^8\,$Pa (e.g., \cite{Bronshten1983}). For iron meteorites, aerodynamic stresses do not necessarily exceed the upper limits of these values, provided the initial velocity is slow enough. This allows for the efficient aerobraking of small impactors and the formation of small impact craters, seen most clearly in Figure~\ref{fig:comet_crater_production_constraints} for iron impactors. This is consistent with the formation of small, young impact craters on the Earth from iron meteorites \cite{DOrazio2011}.
The formation of $100\,$m craters will however be relatively rare, with the majority of impactors being chondritic in composition (e.g., \cite{Walsh2011, Joiret2023}), and typical impact velocities in excess of the Earth's escape velocity (see \S\ref{sec:discussion} for a more detailed discussion of typical crater diameters).
\begin{figure}[t!]
    \centering
    \includegraphics[width=0.75 \textwidth]{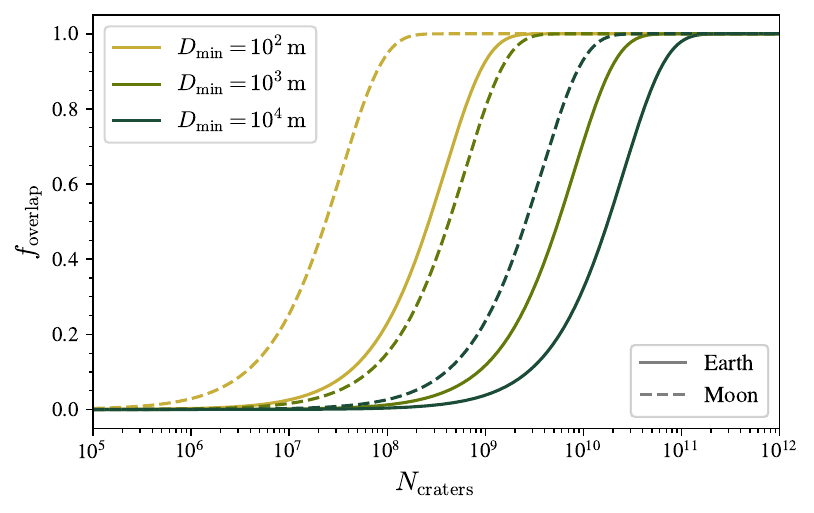}
    \caption{The fraction of overlapping craters on the Earth (solid) and Moon (dashed) is plotted as a function of the number of craters on the surface. We assume a minimum crater diameter of 100\,m (see \S\ref{sec:methods_montecarlo}), and vary the minimum diameter of cometary crater, $D_{\rm min}$, between 100\,m and 10,000\,m.}
    \label{fig:f_overlap_D_min_basic}
\end{figure}

In summary, we find that comets are unable to directly form craters smaller than roughly 1\,km in diameter given the Earth's current atmosphere. Impactors with stone and iron-rich compositions are far more robust to the stresses of atmospheric entry, and can therefore contribute to a population of sub-kilometre craters. In our Monte Carlo simulation (see \S\ref{sec:methods_montecarlo}) we therefore assume a minimum crater size of 100\,m (corresponding to stony and iron-rich impactors), and a minimum cometary crater size of 1\,km.

\subsection{A Monte Carlo model to determine how often craters overlap}
\label{sec:results_montecarlo}
The fraction of overlapping craters on the surface of the Earth, and Moon, is calculated as a function of the number of craters using the Monte Carlo model of \S\ref{sec:methods_montecarlo}, which is shown in Figure~\ref{fig:f_overlap_D_min_basic}. We assume a minimum crater diameter of 100\,m (from the second impact), and vary the minimum diameter, $D_{\rm min}$, of the initial cometary crater between 100\,m and 10,000\,m. Increasing this minimum diameter, $D_{\rm min}$, will decrease the number of cometary craters, as determined by the power-law crater size distribution, and therefore also the fraction of overlapping craters. 

The fraction of overlapping craters on the Moon is also included as dashed lines in Figure~\ref{fig:f_overlap_D_min_basic}. Fewer craters are required for there to be overlapping craters, which differs by a factor of approx.\ $(R_E/R_M)^2$, as would be expected given this is a geometric effect that will scale with surface area. We note that, in the absence of gravitational focusing, the accretional cross-section will also scale as $(R_E/R_M)^2$, and so the fraction of overlapping craters on the surface of the Moon will be similar to on the Earth. 
\begin{figure}[t!]
    \centering
    \includegraphics[width=0.75 \textwidth]{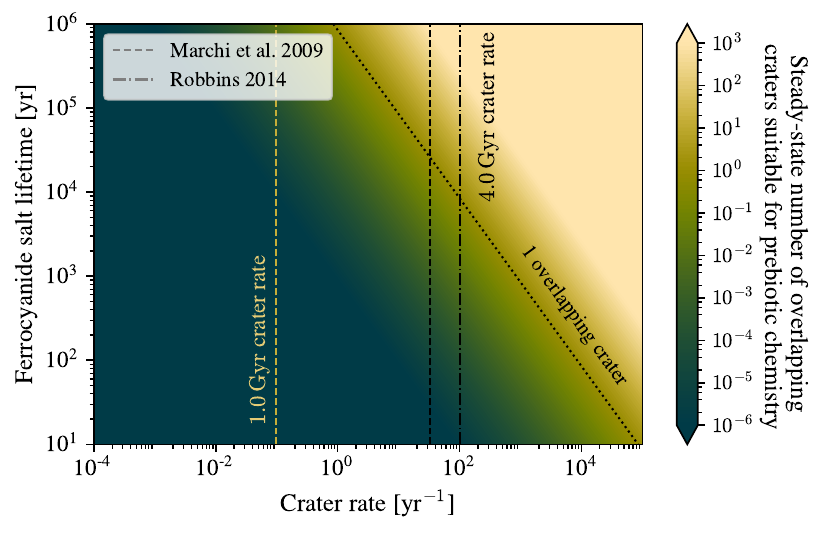}
    \caption{The number of overlapping craters suitable for prebiotic chemistry on the surface of the Earth, in steady-state, is plotted as a function of the ferrocyanide salt lifetime and incident crater rate. The vertical dashed lines indicate the impact rate on the Earth $1$ and $4.0\,$Gyr ago \cite{Marchi2009, Robbins2014}. We assume a minimum cometary crater size of $1\,$km (see \S\ref{sec:results_cratering}), and that comets constitute $1$\% of the impacting population (see \S\ref{sec:methods_montecarlo}). There are no constraints on the second impactor; this could be either cometary or asteroidal.}
    \label{fig:ferro_lifetime_crater_rate_contour}
\end{figure}

As discussed in \S\ref{sec:methods_montecarlo}, there is a competition between the incident impact rate, $R_{\rm in}$, and the typical degradation lifetime of ferrocyanide salts in these craters, $\tau_{{\rm Fe(CN)}_6}$. In steady-state, the number of craters available for prebiotic chemistry is simply $N = R_{\rm in}\tau_{{\rm Fe(CN)}_6}$, and the number of overlapping craters is therefore given by
\begin{equation}
    N_{\rm overlap} = R_{\rm in}\tau_{{\rm Fe(CN)}_6} \cdot f_{\rm overlap}(R_{\rm in}\tau_{{\rm Fe(CN)}_6}),
\end{equation}
where $f_{\rm overlap}(N)$ is the fraction of overlapping craters, as calculated in the Monte Carlo model. In Figure~\ref{fig:ferro_lifetime_crater_rate_contour} we show how the number of overlapping craters (suitable for subsequent prebiotic chemistry), in steady state, depends on both the incident crater rate, and the lifetime of ferrocyanide salts.
We overlay the incident crater rate on the Earth, which suggests that if double impact scenarios are to succeed, they must occur at very early times, and will require the long-term survival of ferrocyanide salts. 

\begin{figure}[t!]
    \centering
    \includegraphics[width=0.75 \textwidth]{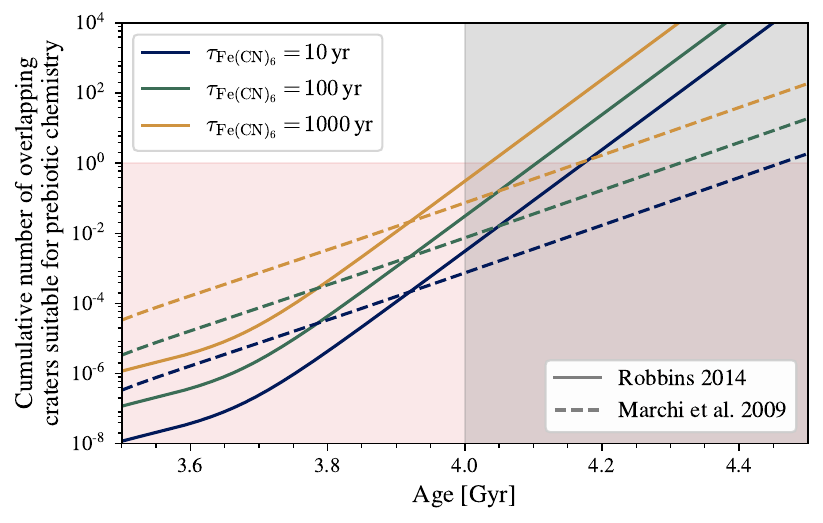}
    \caption{The cumulative number of overlapping craters on the surface of the early Earth is plotted as a function of time, calculated using the impact chronologies of Marchi \cite{Marchi2009} (solid) and Robbins \cite{Robbins2014} (dashed). The impact rate on the early Earth is however uncertain before 4.0\,Gya, which is indicated in grey. Double impact scenarios will not be possible when the cumulative number of overlapping craters is less than 1, which is indicated in red. We require a minimum cometary crater size of 1\,km in the Monte Carlo simulation (see \S\ref{sec:methods_cratering}), and vary the ferrocyanide degradation timescale between 10 and 1000\,yrs. Assuming shallow oceans on the Hadean Earth, we optimistically assume a constant subaerial landmass of 20\%. Finally, it is assumed that comets constitute 1\% of the impacting population, and that 1\% of cometary impacts will successfully deliver prebiotic molecules (see \S\ref{sec:methods_montecarlo}), noting that the number of overlapping craters will scale directly with these fractions.}
    \label{fig:cumulative_overlapping_craters}
\end{figure}

To better constrain the feasibility of double impact scenarios, we calculate the cumulative number of overlapping craters on the early Earth, using the impact chronologies of Marchi et al.\ \cite{Marchi2009} and Robbins \cite{Robbins2014}, as described in \S\ref{sec:methods_chronology}. The results of this calculation are shown in Figure~\ref{fig:cumulative_overlapping_craters}, in which it is clear that even in this best-case scenario, double impact events are very unlikely to occur before 4.0\,Gya, unless the typical lifetime of ferrocyanide salts significantly exceeds 1000\,yrs. As previously discussed in \S\ref{sec:methods_chronology} we should be cautious when extrapolating these chronologies beyond approx.\ 4.0\,Gya, however it is clear that double impact scenarios will require a steep increase in the impact rate, such as with the Robbins \cite{Robbins2014} chronology. The probability, however, that the required reactions for prebiotic syntheses in these environments will occur in the correct sequence, and succeed is of course low \cite{Walton2022, Rimmer2023}. Overlapping craters will therefore be a more attractive setting for the origins of life provided there were a relatively large number of double impact events on the early Earth. This will only be possible at early times ($>4.0\,$Gyr), requiring both the long-term survival of ferrocyanide salts ($>1000\,$yrs), and a very high impact rate.

\section{Discussion}
\label{sec:discussion}

\subsection{The frequency of overlapping craters on the early Earth}

As seen in Figure~\ref{fig:cumulative_overlapping_craters}, it is only at early times at which the impact flux is high enough for their to be any overlapping craters, suitable for prebiotic chemistry, on the surface of the Earth. The number of overlapping craters increases with the typical lifetime of ferrocyanide salts, reflecting the competition between the incident impact rate, and the degradation of ferrocyanide salts, which ultimately governs the ability of a second impactor to successfully reactivate prebiotic chemistry. It is therefore clear that, even in the most optimistic case, double impact scenarios will be constrained to very early times (i.e., $>4.0\,$Gyr), and will require the long-term survival of ferrocyanide salts (i.e., $\gtrsim1000\,$yrs). This timescale has not been measured to the best of our knowledge, however we note this is far longer than the degradation timescale of aqueous ferrocyanide (subject to UV-irradiation), which was found to be (at most) on the order of hours in Todd et al.\ \cite{Todd2022}. This work therefore highlights the importance of measuring this lifetime, possible by heating dry ferrocyanide salts, as this will determine the number of potential environments in which multiple impactors are able to drive prebiotic chemistry.

To numerically model the atmospheric entry of cometary impactors, the resulting impact crater formation, and finally calculate the number of overlapping craters on the surface of the Earth, we necessarily make a number of simplifying assumptions. These assumptions were made, in each case, to maximise the potential number of overlapping craters on the early Earth, and so the true number of overlapping craters will likely be significantly lower than presented in Figure~\ref{fig:cumulative_overlapping_craters}. Whilst the increased impact rates at early times may mitigate some of these assumptions, the intense bombardment of the Hadean Earth will not necessarily be conducive to the emergence, and development of life. Indeed, massive (asteroidal) impactors may have repeatedly sterilised the surface of emergent life \cite{AbramovMojzsis2009, Abramov2013}, with comet-ponds particularly exposed settings for abiogenesis, sensitive to global climatic conditions on the early Earth \cite{MaherStevenson1988}. We therefore argue the typical lifetime of ferrocyanide salts is the more important factor in determining the number of overlapping craters on the early Earth, and therefore the plausibility of double impact scenarios.

The ability of double impact scenarios to provide a plausible setting for the initial stages of prebiotic chemistry will rely on a relatively large number of overlapping craters on the early Earth, to account for the low probability that the multiple stages of prebiotic syntheses will succeed in the correct order \cite{Walton2022, Rimmer2023}. Double impact scenarios are therefore extremely unlikely to occur in the correct conditions for the formation (via the delivery of HCN), and transformation of ferrocyanide, unless the typical lifetime of ferrocyanide salts are in excess of 1000 yrs, and there was a particularly high impact rate on the early Earth (i.e., the Robbins \cite{Robbins2014} impact chronology, as seen in Figure~\ref{fig:cumulative_overlapping_craters}). In all cases, double impact scenarios will be constrained to very early times (before 4.0\,Gya), at which point we reiterate that the impact rate is largely unconstrained (e.g., \cite{StofflerRyder2001}).

\subsection{Model limitations}
\label{sec:discussion_limitations}

The ability of double impact scenarios to drive prebiotic chemistry on the early Earth is contingent on the successful delivery of prebiotic molecules, during both atmospheric entry and impact crater formation, and the environmental conditions on the Hadean Earth. The conclusion that double impact scenarios are possible only at very early times ($>\,$4.0\,Gya) is most sensitive to environmental conditions on the Hadean Earth, with deep planet-wide oceans capable of preventing successful cometary delivery. Similarly, uncertainty in the atmospheric surface density during the Hadean has the potential to rule out the proposed scenario. A significant increase in atmospheric surface density would increase the characteristic rate of cometary fragmentation and thus, markedly increase the minimum cometary diameter able to reach Earth's surface. Given the low tensile strength of comets \cite{Groussin2018}, atmospheric fragmentation is unavoidable. Model results are therefore most sensitive to the bulk density of cometary nuclei, rather than their exact mechanical properties. Any increase in the minimum cometary diameter able to survive atmospheric entry will reduce the total number of successful cometary impacts, and therefore also the number of overlapping craters suitable for prebiotic chemistry. In the following, we discuss in more detail the key factors that would change the conclusions of this work.

\subsubsection{Environmental conditions on the early Earth}

These results are critically sensitive to environmental conditions on the Hadean Earth, with the successful cometary delivery of prebiotic molecules dependent on both the atmospheric density, which will determine the ram pressure experienced by a comet, and the surface area of exposed land, with cometary impacts unable to deliver prebiotically relevant concentrations of HCN to the oceans \cite{Todd2020}. Any increase in atmospheric surface density has the potential to significantly increase the minimum required cometary diameter, as this will lead to the earlier onset of fragmentation, whilst also decreasing the typical timescale of fragmentation (see equation~\ref{eq:comet_deformation}). There are however a lack of well-preserved rocks, and so the atmospheric surface pressure, and extent of oceanic coverage during the Hadean Earth remain subject to debate \cite{CatlingZahnle2020, Charnay2020, Korenaga2021}. We likely overestimate the subaerial landmass during the Hadean, assuming a constant exposed fraction of 20\% \cite{GuoKorenaga2020}, given lower limits restrict the subaerial landmass to hot-spot volcanic islands \cite{BadaKorenaga2018, RosasKorenaga2021}. We note this discussion is general; prospects for successful cometary delivery are greatly reduced, regardless of specific origins scenario, if there were indeed deep oceans, or high atmospheric surface pressures on the Hadean Earth.

\subsubsection{The atmospheric entry of comets}

The fragmentation of comets during atmospheric entry is a complex process, thought to dominate over deceleration and ablation \cite{SvetsovNemtchinov1995, Popova2019}, for which there are several semi-analytical models (e.g., \cite{ArtemievaShuvalov2016}). Our results compare well to Hills \& Goda \cite{HillsGoda1993}, who use a more explicit fragmentation model, and find a minimum cometary diameter of approx.\ 200\,m. This suggests a robustness of our results to the choice of numerical model, however a more detailed 3D model of cometary fragmentation would be required to track the fate of individual fragments, albeit at much larger computational expense.

Cometary survival is however very sensitive to the physical properties of comets, specifically their density and tensile strength, which will determine when, and if fragmentation will occur in the atmosphere. Whilst the tensile strength of comets is poorly constrained, even the strongest comets ($\sim\,$$10^4\,$Pa \cite{Mellor1974, Petrovic2003, Groussin2018}) will still, unavoidably, fragment high in the atmosphere. Comets are likely much weaker, as revealed by Rosetta Comet 67P/Churyumov–Gerasimenko \cite{Kofman2015, Patzold2019}, and more recently the New Horizons fly-by of the Kuiper Belt Object (486958) Arrokoth \cite{Spencer2020, Keane2022}, which support tensile strengths as low as 1\,Pa \cite{Attree2018}. Thus, it is instead the bulk density of cometary nuclei that dominates uncertainty on our results. We assume an upper limit of $0.6\,{\rm g\,cm}^{-3}$ \cite{LowryWeissman2003, Kokotanekova2017}, however note that there are many well-determined measurements in the range $0.1$-$0.4\,{\rm g\,cm}^{-3}$ \cite{AHearn2011, Sierks2015}. These lower values will reduce the typical fragmentation timescale (see equation~\ref{eq:comet_deformation}), therefore increasing the minimum cometary diameter required to survive atmospheric entry. A lower bulk density of only $0.1\,{\rm g\,cm}^{-3}$ would increase the minimum cometary diameter to $\sim\,$1\,km, which is in agreement with simple qualitative considerations (see equation~\ref{eq:r_crit_qualitative}). This would reduce the number of double impact scenarios by approximately one order of magnitude.

\subsubsection{The size distribution of craters on the early Earth}

The results of our Monte Carlo overlap model are sensitive to the choice of power-law SFD, as this will directly determine the number of large (e.g., $>\,$1\,km) craters on the early Earth. For example, a steeper size distribution (i.e $\alpha>2$) would decrease the total number of 1\,km craters. This would imply there were proportionally more small bodies impacting the early Earth, with these smaller comets still unable to reach the surface intact (instead catastrophically fragmenting in the atmosphere). This decrease in the number of successful cometary impacts will correspondingly reduce the number of double impact scenarios on the early Earth. Sensitivity tests reveal that a decrease of power-law index by 0.5 would increase the number of overlapping craters by one order of magnitude (and vice versa). We stress, however, that evidence from the lunar crater record suggests the crater size distribution has remained constant with time \cite{NeukumIvanov1994, Marchi2009, LeFeuvreWieczorek2011, Robbins2014}, and so a large change in power-law index is not expected. This effect is therefore far outweighed by the large uncertainty in impact rate before 4.0\,Gya.

The typical crater diameter from cometary impactors will be significantly larger than the minimum diameter, as calculated in \S\ref{sec:results_cratering}, which corresponds to the smallest impactor capable of surviving atmospheric entry, and an initial velocity of $11.2\,{\rm km\,s}^{-1}$. As discussed in \S\ref{sec:methods_montecarlo}, the mean impact velocity of comets on the Earth is likely larger than $20\,{\rm km\,s}^{-1}$, and perhaps more significantly, the debiased long-period comet SFD (as determined via the Pan-STARSS1 survey) indicates a sharp drop in the number of $<1\,$km comets \cite{Boe2019}. If this is indeed evidence for the non-existence of sub-km comets, rather than a consequence of the inactivity of small comets, this would dramatically increase typical cometary crater diameters. This will accordingly reduce the number of cometary craters (as determined by the power-law SFD), and therefore the number of overlapping craters on the surface of the Earth (see Figure~\ref{fig:f_overlap_D_min_basic}).

\subsubsection{The transformation of ferrocyanide salts by a secondary impactor}

We assume in our Monte Carlo model that a secondary impactor is able to drive subsequent prebiotic chemistry provided it is smaller than the initial cometary impactor, to prevent the destruction of ferrocyanide salts. This requirement may not however be strict enough, with peak pressures achieved during the initial compression stage of crater formation typically in the range 100-1000\,GPa \cite{Melosh1989}. Corresponding temperatures are on the order of several thousand degrees, which in many instances exceeds the melting temperature of rock, and will therefore vaporise any ferrocyanide salts. Shock temperatures will however attenuate rapidly moving away from the point of impact, limiting the destruction of these salts to a relatively localised region around this point. Peak temperatures will therefore be limited to approx.\ 700\,$^\circ$C (as required for the thermal metamorphosis of ferrocyanide salts, e.g., \cite{Patel2015}) across significant regions of the initial cometary crater. Alternatively, heating to temperatures of 700\,$^\circ$C is possible via the slower, sub-escape velocity, collisions of smaller impactors (or fragments thereof) that have been decelerated by the atmosphere, preventing the vaporisation of ferrocyanide salts even at the point of impact\footnote{From equation~\ref{eq:comet_dvdt} we find that an impactor's terminal velocity is given by $v_{\rm term}\approx\sqrt{mg/(C_D\rho_{\rm atm}A)}\propto r^{1/2}$. Low velocity impacts are therefore possible from the fragments of smaller impactors, however as discussed in \S\ref{sec:methods_atmos}, the detailed modelling of atmospheric fragmentation is very challenging. Further work would therefore be necessary in order to accurately model the deposition of meteoritic fragments in small, subaerial ponds.}. These smaller meteorites, decelerated to their terminal velocity, will likely survive impact, at which point they may support more complex prebiotic chemistry through the delivery of catalytic power \cite{Ferris2005, Rotelli2016}, or even the accumulation of nucleobases \cite{Pearce2017}.

\subsection{Plausible routes for prebiotic chemistry}

There are, as we have now discussed, several independent lines of reasoning to suggest that we have (perhaps significantly) overestimated the number of overlapping craters on the early Earth. In light of this, double impact scenarios will be constrained to the Hadean Earth, requiring the long-term survival of ferrocyanide salts (in excess of 1000\,yrs). If the lifetimes of ferrocyanide salts are significantly less than this, or if there was widespread oceanic coverage on the Hadean Earth \cite{Korenaga2021}, double impact scenarios will be extremely unlikely to occur in the correct conditions for the delivery, and transformation of ferrocyanide salts. 

Scenarios invoking the thermal metamorphosis of ferrocyanide salts (important for the release of hydrogen cyanide, cyanamide and cyanoacetylene from ferrocyanide salts; \cite{Patel2015}) do not, however, necessarily require two distinct impacts. Indeed, the heating of ferrocyanide salts could instead be attributed to geothermal heating \cite{Patel2015, Sasselov2020}, removing the requirement for a secondary impactor altogether. This may come in the form of either intrusive, or extrusive igneous activity \cite{Sasselov2020}, and has the benefit of removing the strict (geometric) requirement for high impact rates. We note a plausible landscape of the Hadean Earth could be a series of crater strewn volcanic landmasses, which may more naturally be predisposed to geothermal heating. 
It is imperative therefore to apply experiment, kinetics and equilibrium models of thermal heating of ferrocyanide salts under different redox conditions, to determine whether the subsequent cyanamide, carbide and cyanide salts are indeed formed, with what yield, and with what other salts and minerals.

Alternatively, an initial cometary impact may not be required, given the number of ways to synthesise HCN in a reducing atmosphere (e.g., \cite{Ferus2017a, Parkos2018, Zahnle1986, Zahnle2020, Tian2011, ChameidesWalker1981, Barth2023}). Gaseous HCN is able to dissolve in rainwater, allowing for the accumulation of HCN in shallow subaerial basins and the subsequent formation of ferrocyanide \cite{TonerCatling2019, Sasselov2020}, whilst still allowing for the shock heating of ferrocyanide salts. This appears an attractive proposition, as whilst there exist other ways to produce cyanamide (via the photoredox cycling of thiourea \cite{Liu2020}), and cyanoacetylene (from the degassing of graphite-saturated magmas into hydrothermal systems \cite{ShorttleRimmer2024}), the impact-shock heating of ferrocyanide salts remains a simple, and robust source of both species. Therefore, given the numerous challenges associated with double impact scenarios as outlined in this work, we suggest that the concentration of atmospheric HCN may be a more plausible source of this key feedstock molecule. Given the number of ponds and lakes on the modern Earth (approx. $3\times10^8$ \cite{Downing2006}), models suggest it is likely there will have been a large number of ponds on the Hadean Earth, even when accounting for the growth of subaerial continental crust \cite{Pearce2017}. These ponds are very well-suited for the concentration of ferrocyanide salts, and it is therefore likely that a single impact (either comet or asteroid), capable of transforming ferrocyanide salts, will be possible throughout the Hadean given the high impact rate, and large number of targets.

The efficiency of atmospheric HCN synthesis is, however, dependent on the oxidation state of the early Earth, with many proposed mechanisms requiring a transiently reducing atmosphere \cite{RimmerRugheimer2019}. Recent studies have, however, demonstrated that such conditions may be an expected consequence of late accretion, with the iron-rich cores of large, differentiated impactors readily reducing seawater to generate global, transiently reducing atmospheres on the early Earth \cite{Zahnle2020, Itcovitz2022, Wogan2023}. A `late veneer' of chondritic material ($\sim\,$0.3-0.7\,wt\%\,$M_E$), in the form of large planetesimals, is independently invoked as the simplest explanation for the significant excess of highly-siderophile elements observed in Earth's mantle (e.g., \cite{Walker2009, Bottke2010}),  lending further support to the existence of transiently reducing conditions on the early Earth. The synthesis of HCN is therefore likely both during \cite{Ferus2017a}, and following asteroidal impacts \cite{Zahnle2020, Wogan2023}. We note these synthesis pathways have far less stringent constraints on impact velocity, angle, and impactor diameter than is the case for the successful cometary delivery of HCN \cite{Todd2020}. Whilst further work is required to constrain total HCN synthesis from a purely asteroidal impact flux (and model its subsequent concentration in localised environments), it is possible this could remove any requirement for an initial cometary impact, which otherwise severely limits the number of overlapping craters on the early Earth (see \S\ref{sec:results_montecarlo}).

Notwithstanding the necessity of cometary impacts for the delivery of HCN, as far as we know, they may be the only only source of prebiotic concentrations of cyanoacetylene \cite{MummaCharnley2011}.  Assuming cyanoacetylene survives delivery \cite{Todd2020}, it will need to be used almost immediately, an uninviting prospect considering the importance of access to ultraviolet light, and the importance of temporal and spatial segregation of cyanoacetylene-related chemistry from the initial cyanide-related chemistry \cite{Patel2015}. Instead, it seems cyanoacetylene will need to be stored: converted into a molecule that acts as a kinetically stable compound out of equilibrium, and decomposes into cyanoacetylene upon activation. If this activation is thermal, it may well be that the release of cyanoacetylene will require a subsequent impact. This second impact would be necessary regardless of any alternative sources of cyanide. Thus, if cyanoacetylene is needed for prebiotic chemistry (e.g., \cite{Powner2009, Patel2015, Benner2019, Becker2019}), and comets are the only plausible source of cyanoacetylene, the Monte Carlo model described in this study would be similarly well-suited to quantitatively evaluate the likelihood of these scenarios for the origin of life on Earth.

Generalising these results, the ability of small impactors to deliver feedstock molecules, and drive subsequent prebiotic chemistry on a (exo)planet is dependent on the physical composition of comets, the planet's atmospheric and surface conditions, and the overall impact rate. As such, the potential importance of small impactors, in the context of the origins of life, will vary significantly even within planetary systems. The orbital location of a planet will affect both the overall impact rate, and the impactors' velocity distribution (e.g., Anslow et al.\ \cite{Anslow2023}), whilst the depth of a planet's atmosphere, and the available subaerial landmass will dictate the frequency of successful small cometary impacts. Within the solar system, Mars and Venus provide evidence of this. The cometary delivery of prebiotic molecules will be significantly more successful onto Mars, with mean impact velocities of approx.\ 11.5\,km\,s$^{-1}$, and 24.7\,km\,s$^{-1}$ for Mars, and Venus respectively \cite{JeongAhnMalhotra2015, Greenstreet2012}. Furthermore, the thin Martian atmosphere, and rapid crust formation \cite{Bouvier2018}, in comparison to the thick Venusian atmosphere, should allow for an increased flux of small impactors to the surface, a necessary condition for double impact scenarios.

\section{Conclusion}
\label{sec:conclusions}

This paper investigates the feasibility of double impact scenarios, in which prebiotic feedstock molecules, delivered by an initial cometary impactor, are concentrated via the formation of long-lived ferrocyanide salts, and then thermally transformed into carbide, cyanamide, and cyanide salts via a smaller secondary impactor. This is a very attractive scenario, as these salts will store, segregate and release key feedstock molecules needed for the subsequent synthesis of amino acids, nucleotides and phospholipids. To quantify the number of overlapping craters on the early Earth that are suitable for prebiotic chemistry, we consider both the atmospheric entry of comets, and subsequent impact crater formation.
We find that the role of double impact scenarios as a mechanism for generating favourable environments for the initial stages of prebiotic chemistry is only possible at early times in the Earth’s history ($>$4\,Ga), and if the lifetime of ferrocyanide salts is $>$1000\,yrs.

The cumulative number of overlapping craters on the surface of the Earth, even in the best-case scenario, is significantly less than 1 over a large portion of the Earth's history (0-4\,Gyr). This is a direct consequence of the low impact rate on the Earth and will, even given the long-term stability of ferrocyanide salts, preclude double impact scenarios from providing local conditions conducive to prebiotic chemistry. 
The impact rate before 4\,Gyr is not constrained by the Lunar crater record, however must increase significantly for there to be even a small population of overlapping craters. Double impact scenarios will therefore be possible only on the Hadean Earth, and will require both a very high impact rate (e.g., Robbins, 2014 \cite{Robbins2014}), and the typical lifetime of ferrocyanide salts to be (significantly) in excess of 1000\,yrs. This is of course contingent on the environmental conditions on the Hadean Earth. A dense atmosphere, or widespread oceanic coverage would both markedly decrease the number of overlapping craters on the early Earth, rendering double impact scenarios implausible settings for the initial stages of prebiotic chemistry.

If the lifetime of ferrocyanide salts is less than 1000\,yrs (as is the case for aqueous ferrocyanide; \cite{Todd2022}), scenarios requiring secondary impacts will be much more likely to succeed via the initial concentration of atmospheric HCN, as opposed to external cometary delivery. Conversely, if the cometary delivery of prebiotic molecules is indeed important for the initial concentration of prebiotic molecules, then the external heating of reactants must instead come from external (e.g., geothermal) sources, rather than additional impacts. 
The work emphasises the need to measure the typical lifetime of ferrocyanide salts in early Earth environments, which will ultimately determine the relative importance of double impact scenarios in delivering prebiotic feedstock molecules, and driving subsequent prebiotic chemistry. 
More generally, these results highlight how the global environmental conditions on a planet, particularly the surface atmospheric density, and the extent of subaerial landmass, will constrain the successful cometary delivery of prebiotic molecules, independent of the specific origins scenario.

\medskip

\noindent \textbf{Data accessibility.} The code (and numerical models) used to generate all figures in the manuscript can be found in the following repository: \url{https://github.com/richard17a/overlap}.

\medskip

\noindent \textbf{Acknowledgments} 
We thank two anonymous reviewers for insightful comments that have greatly improved this manuscript.
R.J.A. acknowledges the Science and Technology Facilities Council (STFC) for a PhD studentship. R.J.A thanks Greg Cooke, Claire Guimond, Zoe Todd, and Robin Wordsworth for helpful discussions. A.B. acknowledges the support of a Royal Society University Research Fellowship, URF/R1/211421. A.S.P.R. acknowledges funding from Trinity College Cambridge. C.H.M. gratefully acknowledges support from the Leverhulme Centre for Life in the Universe at the University of Cambridge. C.R.W. acknowledges funding from the NOMIS Foundation [ETH-NOMIS COPL fellowship] and Trinity College.\\
\textit{Software acknowledgements:} \textsc{numpy} \cite{Harris2020}, \textsc{numba} \cite{Lam2015}, \textsc{scipy} \cite{Virtanen2020}, \textsc{matplotlib} \cite{Hunter2007}

%%%%%%%%%% Insert bibliography here %%%%%%%%%%%%%%

\vskip2pc

\bibliographystyle{RS}
\bibliography{main}

\end{document}